\documentclass[twocolumn,showpacs,preprintnumbers,aps,amsmath]{revtex4}
\usepackage{amsmath,amssymb}
\usepackage{latexsym}
\usepackage[dvips]{graphicx}

\begin{document}

\preprint{EPHOU 08-002 }

\title{Moduli fixing and T-duality in Type II brane gas models}

\author{Masakazu Sano and Hisao Suzuki}

\affiliation{ Department of Physics,  Hokkaido University, Sapporo, Hokkaido 060-0810 Japan}

\begin{abstract}
 
We consider a compactification with a six-dimensional torus in the type II brane gas models. 
We show that the dilaton and the scale of each cycle of the internal space are fixed in the presence of NS5-branes and Kaluza-Klein monopoles as well as D-branes with the gauge fields. We can construct various models that lead to fixed moduli by using T-duality transformations. 
\end {abstract}

\pacs {04.50. -h, 11.25.Mj, 11.25. -w}

\maketitle

\section {Introduction}
Brane gas models have been investigated to understand the origin of the hierarchy of dimensions. Since superstring theories predict 
ten-dimensional universe, we need to know the reason why four-dimensional space-time expands whereas the other spaces are small if the theories are the correct description of our space-time.  This problem is related to the origin of our universe and it is natural to consider this problem in the framework of the cosmology.
Brandenberger and Vafa \cite{bio_BV} considered momentum and winding modes of strings which 
wrap around a 9-dimensional space.
It was argued that the ($3+1$)-dimensional space-time could 
expand, so that such a space-time loses its winding of strings.  
This idea was polished up by Tseytlin and Vafa \cite{bio_TV} in the context of dilaton gravity. 
A extension including the effect of a gas of D$p$-branes has been analyzed by 
Alexander, Brandenberger and Easson \cite{bio_ABE}. 
The T-duality in the brane gas model was investigated in Ref.\cite{bio_BB}. 
The cosmological evolution of String/Brane gas models was considered in 
\cite{bio_EGJK1, bio_EGJK2, bio_EGJK3, bio_KW}

One of the problems of the brane gas models is the stability of the moduli parameters.  A lot of investigations concerning with this problem
have been performed \cite{bio_WB,bio_KR,bio_BBST1,bio_BW,bio_W2,bio_BP1,bio_BP2, bio_P2, bio_KS, bio_CWB, bio_C1, bio_BC, bio_BBC, bio_ET}.
It was expected that winding modes of D$p$-branes prevent the internal space from expanding, 
because D$p$-branes give negative pressure in the internal space \cite{bio_ABE}. 
If the dilaton field is fixed, the internal space becomes stable, 
using the effect of winding modes of D$p$-branes \cite{bio_KR}. 
On the other hand,  we have to consider the dilaton field at the early stage of the universe.
However a simple brane gas model with the dilaton field cannot stabilize both 
the internal scale factor and the dilaton simultaneously, 
although the linear combination of them can be stabilized \cite{bio_BBC}. 
In Ref.\cite{bio_CW}, the issue of stabilizing the dilaton was considered, assuming the existence 
of tensionless membranes which require the quantum analysis of supermembrane. 
Recently, Danos, Frey and Brandenberger suggested a moduli fixing procedure in the heterotic strings on an orbifold \cite{bio_DFB}. 
In their models, the gaugino condensation has a crucial role for the moduli fixing.  
Although many works have been done, 
  it still remains to be an unavoidable problem how the dilaton and the scale of the internal space are fixed simultaneously
for the description of  a realistic cosmology in the brane gas model.

In this paper, we will concentrate on the problem of the moduli fixing in the type II brane gas model, 
using Dirac-Born-Infeld action. 
We will show that the NS5-branes or the Kaluza-Klein monopoles (KK5-monopoles) play a crucial role 
for the stability of the dilaton field.
It is known that 
the world volume action of the KK5-monopoles and the NS5-branes is 
derived by the dimensional reduction of the 11-dimensional KK-monopole \cite{bio_BJO} and the M$5$-brane \cite{bio_BLO1} respectively.  
We will analyze the role of the gauge fields on the D$p$-brane and the world volume action of the NS5-brane and the KK5-monopole
 in addition to the D$p$-branes. 
We will take the six-dimensional torus $T^{6}$ as the internal space for simplicity.
If those objects wrap over the cycles of $T^{6}$, we will show that the dilaton and the scale of $T^{6}$ are fixed. 
We can find various D-brane configurations which lead to the fixed moduli spaces and dilaton by using T-dualities. Namely, 
the T-duality allows us to obtain dual type II brane gas models once the moduli fields are fixed.

This paper is organized as follows. 
In the next section, we will prepare the bulk action and the metric defined by the four-dimensional Einstein frame.  
In Sec. III, we discuss the world volume action of the D$p$-branes. 
In Sec. IV,  we will show the construction of the world volume action of the NS5-branes and the KK5-monopoles.
In Sec. V, we will investigate the problems of  the moduli fixing, using the world volume action of the D$p$-branes, NS5-branes and KK5-monopoles. 
Sec. VI will be devoted to the conclusions and some discussions.

\section{Bulk action and four-dimensional Einstein frame}

In this section, we would like to provide the set-up of the bulk action and 
clarify the relation between the string frame and the four-dimensional Einstein frame, 
because the moduli fixing should be considered by the four-dimensional Einstein frame. 

In the present paper, 
we assume that the winding modes of large directions have annihilated and the three-dimensional space is the Euclid space.  
We also assume that D$p$-branes, NS5-branes and KK5-monopoles distribute homogeneously in the internal space. 
Taking into account those assumptions, in the string frame, 
we consider a ten-dimensional metric with a $T^{6}$ compactification as 
\begin{align}
ds^{2}_{10}= -e^{2 \lambda _{0}(t)}dt^{2}+e^{2 \lambda (t)} d\mathbf{x}^{2} +\sum_{m=4}^{9} e^{2 \lambda_{m} (t)} (dy^{m})^{2} 
\label{a1}
\end{align}   
where $d\mathbf{x}^{2}\equiv \sum_{i=1}^{3}(dx^{i})^{2}$ represents a three-dimensional flat space. 
The scale factor $\exp{[\lambda _{m}(t)]}$ is the moduli of $T^{6}$ which represents the size of the torus.
Using the degree of freedom of the scale factor $\exp{[\lambda _{m}(t)]}$, 
we can define $T^{6}$ by $y^{m} \sim  y^{m}+2 \pi \sqrt{\alpha '}$. This construction of $T^{6}$ 
indicates that the radius of the cycle of $T^{6}$ is given by $2\pi R_{m}=\int dy^{m} \exp{[\lambda_{m}(t)]}=
2 \pi \sqrt{\alpha'} \exp{[\lambda _{m}(t)]}$ for each cycle.

In the brane gas model, we will focus on the dilation gravity sector of the bulk action of the closed string: 
\begin{equation}
S_{0}=\frac{1}{16 \pi G_{10}} \int d^{9+1}X \sqrt{-G(X)} e^{-2 \phi} [ R+4(\nabla \phi)^{2} ] \label{a2}
\end{equation}
We assume that the dilaton field depends only on time, namely, $\phi=\phi(t)$. 
It is known that this action is invariant under the T-duality transformation \cite{ bio_TV, bio_BHO} given by
\begin{equation}
\begin{split}
\lambda _{m}(t)\rightarrow -\lambda _{m}(t), \quad
\quad \phi(t) \rightarrow \phi(t)-\lambda _{m}(t).  \label{a3}
\end{split}
\end{equation}

In order to obtain the four-dimensional Einstein-Hilbert term, 
we will consider the following transformation:
\begin{equation}
\begin{split}
\lambda _{0}(t)&=n(t)+\phi(t)-\frac{1}{2}\overline{\lambda }(t) , \\
\lambda (t)&=A(t)+\phi(t)-\frac{1}{2}\overline{\lambda }(t) , \\  
\overline{\lambda }(t)&\equiv \sum_{m=4}^{9}\lambda _{m}(t)
\end{split} \label{a5}
\end{equation}
In these variables,  $\exp{[n(t)]}$ and $\exp{[A(t)]}$ represents the lapse function and the three-dimensional scale factor in the four-dimensional Einstein frame. 
Indeed, the metric (\ref{a1}) implies that the coefficient of $R$ is 
proportional to 
$\exp{[ \lambda_{0}(t)+3\lambda (t)] } \times \exp{[ \sum_{m=4}^{9} \lambda _{m}(t)-2 \phi(t)]} R$  
which can be normalized as
$\exp{[ n(t)+3A(t)} ] R_{4}+\cdots=\sqrt{-g_{4}(x)}R_{4}+\cdots $ under the transformation (\ref{a5}). 

\section{World volume action of D-brane}

In order to construct the brane gas model, we have to include 
the world volume action of D$p$-brane. 
The role of the winding mode of the D-brane has been investigated, 
however the role of the gauge fields on the D-brane is still unclear in the brane gas model. 
In this section, we would like to derive the potential of the world volume action of the D-brane with the gauge fields,   
because the explicit form of the potential is required to discuss the moduli fixing.   

We consider the D$p$-brane wrapping over the cycles of the six-dimensional torus $T^{6}$. 
If the D$p$-brane wraps over the ($m_{1}, \dots ,m_{p}$)-cycle ($0\leq  p\leq 6$),   
the gauge fields on the D$p$-brane exist in the ($m_{1}, \dots ,m_{p}$)-directions.
We assume that the gauge field is Abelian and that the gauge potential and the transverse coordinates depend only on the time variable, i.e. 
$A_{m_{a}}(t)$ and $X^{m_{a}}(t)$ which means the homogeneous distributions of the branes.
We will adopt the coordinate system as $\xi^{0}=t$, $\xi^{m_{a}}=y^{m_{a}}$ ($a=1,\,2,\,\dots ,\,p$).
Then, the Dirac-Born-Infeld  action of D$p$-brane wrapping over the ($m_{1}, \dots ,m_{p}$)-cycle is given by
\begin{align}
&S_{\text{D}p}^{(m_{1}\cdots m_{p})} \notag \\
=&-T_{p}\int_{R\times \varSigma_{p}}d^{p+1}\xi e^{-\phi(t)}\sqrt{-\det(\gamma_{ab}+2\pi \alpha' F_{ab})} \notag \\
=&-(2\pi \sqrt{ \alpha'} \,)^{p}T_{p}\int dt e^{-\phi(t)+\lambda _{0}(t)+ \sum_{a=1}^{p}\lambda _{m_{a}}(t)}  \notag \\
&\times \Bigl\{ 1 -(2\pi \alpha')^{2}\sum_{a=1}^{p}e^{-2 \lambda _{0}(t)-2\lambda _{m_{a}}(t)}(\dot{A}_{m_{a}}(t))^{2} \notag \\
&-\sum_{b=p+1}^{6}e^{-2 \lambda _{0}(t)+2 \lambda _{m_{b}}(t)} (\dot{X}^{m_{b}}(t))^{2} \Bigr\}^{\frac{1}{2}} \label{a7}
\end{align}
where $T_{p}=(2\pi)^{-p}(\alpha')^{-(p+1)/2}$, 
 $\alpha'$ is related with the string length as $l_{s}=\alpha'^{1/2}$
 \cite{bio_P} and $\int_{\varSigma_{p}} d^{p}\xi=(2\pi \sqrt{\alpha'})^{p}$. 

The T-duality of the Dirac-Born-Infeld action of the D$p$-brane was shown in \cite{bio_BR}. 
The analysis is general and applicable to our brane gas model.  
We will consider the T-dual of the ($m_{p}$)-cycle. 
This T-dual requires that the dilaton $\phi(t)$ and the scale factor $\exp{[\lambda _{m_{p}}(t)]}$ transform as (\ref{a3}). 
We also impose the following dual transformations between the gauge fields and the transverse coordinates: 
\begin{equation}
2\pi \alpha'  A_{m_{p}}(t)\rightarrow X^{m_{p}}(t). \label{a8}
\end{equation}
Using those T-dual transformations (\ref{a3}) and (\ref{a8}) as well as the relation of the tension
$(2\pi\sqrt{\alpha'})^{p}T_{p}=(2\pi\sqrt{\alpha'})^{p-1}T_{p-1}$, we can see that the D$p$-brane action 
$S_{\text{D}p}^{(m_{1}\cdots m_{p})}$ is mapped to 
the D$(p-1)$-brane action $S_{\text{D}(p-1)}^{(m_{1}\cdots m_{p-1})}$ which wraps over the $(m_{1}\cdots m_{p-1})$-cycle. 
We can also consider  the T-dual of a transverse direction, i.e. ($m_{p+1}$)-cycle. 
Applying the T-dual rule (\ref{a3}) for $\lambda_{m_{p+1}}(t)$ and 
\begin{equation}
X^{m_{p+1}}(t) \rightarrow 2\pi \alpha'  A_{m_{p+1}}(t), \label{a9}
\end{equation}
we find that the D$p$-brane action $S_{\text{D}p}^{(m_{1}\cdots m_{p})}$ is mapped to 
the D$(p+1)$-brane action $S_{\text{D}(p+1)}^{(m_{1}\cdots m_{p+1})}$. 

We need to know the explicit form of the potential term in the four-dimensional Einstein frame for investigating the moduli fixing problems.
Substituting (\ref{a5}) for equation (\ref{a7}), we obtain the world volume action in the four-dimensional Einstein frame: 
\begin{align}
&S_{\text{D}p}^{(m_{1}\cdots m_{p})} \notag \\
=&-(2\pi\sqrt{\alpha' } )^{p}T_{p}\int dt e^{n(t)-\frac{1}{2}\overline{\lambda }(t)+ \sum_{a=1}^{p}\lambda _{m_{a}}(t)}  \notag \\
&\times \Bigl\{ 1 -\mathcal{A}(t) \Bigr\}^{\frac{1}{2}} \label{a10}
\end{align}  
where $\mathcal{A}(t)$ is
\begin{align}
\mathcal{A}(t)&\equiv (2\pi \alpha')^{2}\sum_{a=1}^{p}e^{-2\phi(t) -2 n(t)+\overline{\lambda }(t)-2\lambda _{m_{a}}(t)}(\dot{A}_{m_{a}}(t))^{2} \notag \\
&+\sum_{b=p+1}^{6}e^{-2 \phi(t)-2 n(t) +\overline{\lambda }(t)+2\lambda _{m_{b}}(t)} (\dot{X}^{m_{b}}(t))^{2}. \label{a11}
\end{align}

We can solve the equations of motion of $A_{m_{a}}(t)$ and $X^{m_{b}}(t)$ derived from (\ref{a10}) as follows:  
\begin{align}
&e^{n(t)-\frac{1}{2}\overline{\lambda }(t)+ \sum_{a'=1}^{p}\lambda _{m_{a'}}(t)} \notag \\
&\qquad \times (2\pi \alpha') 
e^{-2\phi(t) -2 n(t)+\overline{\lambda  }(t)-2\lambda _{m_{a}}(t)} \dot{A}_{m_{a}}(t) \notag \\
=&(2\pi \alpha')^{-1}f_{m_{a}} |^{\parallel m_{1}\cdots m_{p}}
\Bigl\{ 1 -\mathcal{A}(t) \Bigr\}^{\frac{1}{2}}, \label{a12} \\[15pt]
&e^{n(t)-\frac{1}{2}\overline{\lambda  }(t)+ \sum_{a'=1}^{p}\lambda _{m_{a'}}(t)} \notag \\
&\qquad \times e^{-2\phi(t)-2 n(t) +\overline{\lambda }(t)+2\lambda _{m_{b}}(t)}\dot{X}^{m_{b}}(t) \notag \\
=&v^{m_{b}}|^{\perp m_{1}\cdots m_{p}} 
\Bigl\{ 1 -\mathcal{A}(t) \Bigr\}^{\frac{1}{2}} \label{a13}
\end{align}
where $f_{m_{a}} |^{\parallel m_{1}\cdots m_{p}}$ and 
$v^{m_{b}} |^{\perp m_{1}\cdots m_{p}}$ are constants of integration. 
We can solve equations (\ref{a12}) and (\ref{a13}) on $\mathcal{A}(t)$: 
\begin{equation}
\begin{split}
\mathcal{A}(t)&=\frac{  e^{-2\sum_{a'=1}^{p}\lambda _{m_{a'}}(t)} \widetilde{\mathcal{A}}(t) }
{1+ e^{-2\sum_{a'=1}^{p}\lambda _{m_{a'}}(t)} \widetilde{\mathcal{A}}(t)  }, \\
 \widetilde{\mathcal{A}}(t)&\equiv \sum_{a=1}^{p}e^{2\phi (t)+2 \lambda _{m_{a}}(t)} 
\Bigg\{ \frac{  f_{m_{a}}|^{\parallel m_{1}\cdots m_{p}} }{2\pi \alpha'} \Biggr\}^{2}  \\
&+\sum_{b=p+1}^{6}e^{2\phi(t)-2 \lambda _{m_{b}}(t)} \Biggl\{v^{m_{b}} |^{\perp m_{1}\cdots m_{p}} \Biggr\}^{2} .
\label{a14}
\end{split}
\end{equation}
In the four-dimensional Einstein frame, the potential term is derived by 
\begin{align}
u_{\text{D}p}^{(m_{1}\cdots m_{p})} \equiv-T^{0}_{~0}&=\frac{2 g^{00}}{\sqrt{-g_{4}(x)}}\frac{\delta \mathcal{L}_{\text{D}p}^{(m_{1}\cdots m_{p})}}{\delta g^{00}} \notag \\
&=-e^{-n(t)-3A(t)}\frac{\delta \mathcal{L}_{\text{D}p}^{(m_{1}\cdots m_{p})}}{\delta n(t)} . \label{a15}
\end{align}
Using (\ref{a10}), (\ref{a14}) and (\ref{a15}), 
we obtain the potential term of the D$p$-brane wrapping over the ($m_{1} \cdots m_{p}$)-cycle:
\begin{align}
u_{\text{D}p}^{(m_{1}\cdots m_{p})} 
=& e^{-3A(t)} (2\pi \sqrt{\alpha' } )^{p}T_{p}  \notag \\
&\times \Bigl\{ e^{ -\overline{\lambda }(t)+ 2 \sum_{a=1}^{p}\lambda _{m_{a}}(t)} + e^{-\overline{\lambda }(t)} 
 \widetilde{\mathcal{A}}(t)  \Bigr\}^{\frac{1}{2}}. \label{a16}
\end{align}
Taking into account the definition of 
$ \widetilde{\mathcal{A}}(t) $, it is easily found that $u_{\text{D}p}^{(m_{1}\cdots m_{p})}$ cannot fix the dilaton.

\section{World volume action of NS5-brane and KK5-monopole}

Type IIA/IIB theory has NS5-brane and the KK5-monopole. 
However, the cosmological role of those objects has not been understood well in the brane gas model (NS5-brane was discussed in \cite{bio_BBC}). 
In this section, we will consider the effect of the NS5-brane and the KK5-monopole. 
The T-duality of those objects and the potential will be shown below. 

The world volume action of the static KK5-monopole wrapping over the ($m_{1}\cdots m_{5}$)-cycle is given by \cite{bio_BJO, bio_EJL} 
\begin{align}
&S^{(m_{1}\cdots m_{5})}_{\text{KK}5} 
=-T_{\text{KK}5}\int_{R\times \varSigma_{5}} d^{6}\xi e^{-2\phi(t)} k^{2} \sqrt{-\det \widetilde{\gamma}_{ab}} \notag \\
=&-(2\pi \sqrt{\alpha' } )^{5}T_{\text{KK}5} \int dt e^{-2\phi(t)+2\lambda _{6}(t)+\lambda _{0}(t)+\sum_{a=1}^{5}\lambda _{m_{a}}(t)}, \notag \\
&\widetilde{\gamma}_{ab} \equiv \frac{\partial X^{m}}{\partial \xi^{a}} \frac{\partial X^{m}}{\partial \xi^{a}} 
( G_{mn}-k^{-2}k_{m}k_{n}  ).  
\label{a17} 
\end{align}
where $k^{m}\equiv \delta^{m m_{6}}$ is the killing vector of the $S^{1}$ isometry of the ($m_{6}$)-cycle and 
$k^{2}\equiv G_{mn}k^{m}k^{n}=\exp{[2 \lambda_{m_{6}}(t)]}$.  
We also consider the static NS5-brane. 
The winding mode of the NS5-brane \cite{bio_BLO1} wrapping over ($m_{1}\cdots m_{5}$)-cycles is given by 
\begin{align}
&S^{(m_{1}\cdots m_{5})}_{\text{NS}5}
=-T_{\text{NS}5}\int_{R\times \varSigma_{5}} d^{6}\xi e^{-2\phi(t)} \sqrt{-\det\gamma_{ab}} \notag \\
=&-(2\pi \sqrt{\alpha' })^{5}T_{\text{NS}5} \int dt e^{-2\phi(t)+\lambda _{0}(t)+\sum_{a=1}^{5}\lambda _{m_{a}}(t)}. \label{a18}
\end{align}

The T-dual transformations of the world volume action of the NS5-brane and the KK5-monopole have been shown in \cite{bio_BJO, bio_EJL}.  
The type IIA NS5-brane is mapped to the type IIB NS5-brane by T-duality transformation with respect to a tangent direction. 
For example, if we take the T-dual of $\lambda _{5}(t)$, the NS5 world volume action (\ref{a18}) is mapped to 
the NS5 world volume action under (\ref{a3}). 
Similarly, we are able to check that the type IIA KK5-monopole is mapped to the type IIB KK5-monopole by a T-dual with respect 
to a tangent direction. 
The T-dual to a transverse cycle, on the other hand, gives an exchange of the NS5-brane for the KK5-monopole, if 
$T_{\text{NS}5}=T_{\text{KK}5}$ \cite{bio_EJL}.  
In fact, we can show that the T-dual of the NS5-brane along the ($m_{6}$)-cycle becomes the KK5-monopole under the T-dual rule (\ref{a3}).  

To obtain the potential term of the NS5-brane and the KK5-monopole in the four-dimensional Einstein frame, 
we consider the change of fields as in  (\ref{a5}). 
Substituting (\ref{a5}) for (\ref{a17}) and (\ref{a18}) and replacing $\mathcal{L}_{\text{D}p}^{(m_{1}\cdots m_{p})}$ with $\mathcal{L}_{\text{NS}5/\text{KK}5}^{(m_{1}\cdots m_{5})}$ in 
(\ref{a15}), the potential terms of the NS5-brane and the KK5-monopole are given by 
\begin{align}
u_{\text{NS}5}^{(m_{1}\cdots m_{5})}&=e^{-3A(t)}(2\pi\sqrt{\alpha' } )^{5}T_{\text{NS}5} \notag \\
&\qquad  \times e^{-\phi(t)-\frac{1}{2}\overline{\lambda  }(t)+\sum_{a=1}^{5}\lambda _{m_{a}}(t)}, \label{a20} \\
u_{\text{KK}5}^{(m_{1}\cdots m_{5})}&=e^{-3A(t)}(2\pi\sqrt{\alpha' })^{5}T_{\text{KK}5} \notag \\
&\qquad  \times e^{-\phi(t)-\frac{1}{2}\overline{\lambda  }(t)+2\lambda _{m_{6}}+\sum_{a=1}^{5}\lambda _{m_{a}}(t)}. \label{a21}
\end{align}

\section{Moduli fixing and T-duality}

In this section, we are going to show that there exist models where 
the dilaton field and scales of $T^{6}$ are fixed simultaneously. 
Once moduli fields will be fixed,  
we are able to obtain dual type IIA/IIB brane gas models by the T-duality transformations. 

All we need is the existence of D$p$-branes with gauge fields and NS5-branes or KK5-monopoles. At first, we consider the type IIB theory with D1-branes and KK5-monopoles. We require that these branes and monopoles distribute homogeneously. Namely, we consider a type IIB brane gas model in which the D1-branes wrap over each 1-cycle and 
the KK5-monopoles wrap over the ($45678$)-cycle and its cyclic permutations. 
Although we can see the stability for general initial conditions,
we take $f_{m_{a}}|^{\parallel m_{1}\cdots m_{p}} \equiv 2 \pi f$, $v^{m_{a}}|^{\perp m_{1}\cdots m_{p}}=0$ to obtain the analytic value of $\lambda_{a}(t)$ and $\phi(t)$ at a minimum. 
The choice of this initial condition represents 
the fact where the initial gauge fields is the same for each cycle.
This condition may be natural for the isotropic expansion of the internal space.
The above conditions provide us the following potential term: 
\begin{align}
U^{\text{IIB}} &=\frac{N_{\text{D1}}}{e^{3A(t)}}  (2 \pi \sqrt{\alpha'})T_{\text{1}}\sum_{a=4}^{9}e^{-\frac{1}{2}\overline{\lambda}(t)+\lambda_{a}(t)}(1+f^{2}e^{2\phi(t)})^{\frac{1}{2}}\notag \\
&\qquad   +\frac{N_{\text{KK5}}}{e^{3A(t)}}(2 \pi \sqrt{\alpha'})^{5}T_{\text{KK5}}
\sum_{a=4}^{9} e^{-\phi(t)+\frac{1}{2} \overline{\lambda}(t)+\lambda_{a}(t)}    \label{a22}
\end{align}
where we have used (\ref{a16}) and (\ref{a21}). 
$N_{\text{D1/KK5}}$ denotes the number of the D$1$-branes and the KK5-monopoles 
and $N_{\text{D1/KK5}}e^{-3A(t)}$ is the number density. 
The first term indicates the D$1$-brane action. The second term describes the world volume action of the KK$5$-monopole. 
The minimum is given by $\partial U^{\text{IIB}} /\partial \phi=\partial U^{\text{IIB}} /\lambda_{a} =0$ which derives the following equations: 
\begin{align}
&(2\pi \sqrt{\alpha'})T_{\text{1}}N_{\text{D1}}\sum_{a=4}^{9}e^{-\frac{1}{2}\overline{\lambda}(t)+\lambda_{a}(t)}f^{2}e^{2\phi(t)}(1+f^{2}e^{2\phi(t)})^{-\frac{1}{2}} \notag \\
&-(2\pi \sqrt{\alpha'})^{5}T_{\text{KK5}}N_{\text{KK5}}\sum_{a=4}^{9} e^{-\phi(t)+\frac{1}{2} \overline{\lambda}(t)+\lambda_{a}(t)}=0, \label{a23} \\
&(2\pi \sqrt{\alpha'})T_{\text{1}}N_{\text{D1}}(2e^{\lambda_{a}(t)}-\sum_{b=4}^{9}e^{\lambda_{b}(t)})e^{\phi(t)}(1+f^{2}e^{2\phi(t)})^{\frac{1}{2}} \notag \\
&+(2\pi \sqrt{\alpha'})^{5}T_{\text{KK5}}N_{\text{KK5}}
e^{\overline{\lambda}(t)}(2e^{\lambda_{a}(t)}+\sum_{b=4}^{9}e^{\lambda_{b}(t)}) =0. \label{a24}
\end{align}
Using (\ref{a24}), ($\frac{\partial}{\partial \lambda_{a}}-\frac{\partial}{\partial \lambda_{b}})U^{\text{IIB}}=0$ gives 
\begin{equation}
e^{\lambda_{a}(t)}=e^{\lambda_{b}(t)}  \label{a25}
\end{equation}
where ($a, b=4,\,5,\,\dots ,\,9$).  
We will define $\lambda'(t)=\lambda_{4}(t)=\cdots =\lambda_{9}(t)$. 
Substituting (\ref{a25}) for (\ref{a23}) and (\ref{a24}), the result is given by 
\begin{equation}
e^{2\phi_{\text{min.}}}=\frac{1}{f^{2}}, \qquad 
e^{2 \lambda'_{\text{min.}}}=\Biggl( \frac{N_{\text{D1}}}{N_{\text{KK5}}} \Biggr)^{\frac{1}{3}}\Bigl( \frac{1}{2 f^{2}} \Bigr)^{\frac{1}{6}}  \label{a26}
\end{equation}
where we have used $T_{\text{KK5}}=T_{\text{NS}5}=T_{5}$ and 
$T_{p}=(2\pi)^{-p}(\alpha')^{-(p+1)/2}$. 
If we choose the initial condition of the gauge fields as $1 \ll  f^{2}$, the weak string coupling 
is realized. $N_{\text{D1}}$ and $N_{\text{KK5}}$ control the scale of the six-dimensional torus. 

Therefore, we see that the dilaton and the scale of $T^{6}$ can be fixed in the type IIB brane gas model. 
If $U^{\text{IIB}}$ does not include the gauge fields or the KK$5$-monopoles, 
we cannot fix the dilaton field and the scale of $T^{6}$ simultaneously.  
It is stressed that the scale of $T^{6}$ becomes isotropic because of the isotropic initial conditions for gauge fields on D1-branes.
This property can be seen from the invariance of $U^{\text{IIB}}$ under $\lambda_{a} \leftrightarrow \lambda_{b}$. 
We have argued with one type of D-branes. 
Moreover, we have considered $v^{m_{a}}|^{\perp m_{1}\cdots m_{p}}=0$.
However, adding other branes or $v^{m_{a}}|^{\perp m_{1}\cdots m_{p}}$-terms to (\ref{a22}) does not ruin the existence of the minimum of $U^{\text{IIB}}$, 
because the potential of the D$p$-branes, the NS5-branes and the KK5-branes is positive which does not change the fact that total potential runs to infinity at large value of the moduli fields.

T-duality guarantees the existence of other type of models with the same characteristic. Let us consider the T-duality transformation of the model (\ref{a22}). For example, the T-duality of the ($456789$)-cycle gives rise to the dual type IIB brane gas model 
which includes D5-branes and NS5-branes. 
In fact, the T-dual transformation rules (\ref{a3}), (\ref{a8}) and (\ref{a9}) gives the following potential: 
\begin{align}
&U_{\text{dual}}^{\text{IIB}} \notag \\
=&
\frac{N^{\text{dual}}_{\text{D5}}}{e^{3A(t)}}  (2\pi \sqrt{\alpha'})^{5}T_{\text{5}}\sum_{a=4}^{9}e^{\frac{1}{2}\overline{\lambda}(t)-\lambda_{a}(t)}
(1+f^{2}e^{2\phi(t)-2\overline{\lambda}(t)})^{\frac{1}{2}}\notag \\
&\qquad +\frac{N^{\text{dual}}_{\text{NS5}}}{e^{3A(t)}}  (2\pi \sqrt{\alpha'})^{5}T_{\text{NS5}}\sum_{a=4}^{9} e^{-\phi(t)+\frac{1}{2} \overline{\lambda}(t)-\lambda_{a}(t)} \label{a27}
\end{align}
where $N^{\text{dual}}_{\text{D5}}\equiv N_{\text{D1}}$ and $N^{\text{dual}}_{\text{NS5}}\equiv N_{\text{KK5}}$.
The first term indicates the D$5$-brane action moving to the transverse direction. 
The second term describes the world volume action of the NS5-branes. 
$e^{2\phi_{\text{min.}}}$ and $e^{2 \lambda'_{\text{min.}}}$ are mapped to the dual values 
which are not self-dual by (\ref{a3}). 

Note that we can obtain possible dual type IIA or IIB models by the T-duality.

\section{Conclusion and discussions}

In this paper, we have considered the moduli fixing problem in the type II brane gas model.  
It is stressed that this moduli fixing including dilaton field is realized in the presence of NS5-branes or KK5 -monopoles as well as the winding modes of the D-brane and the effects of the gauge fields on the D-brane. 
NS5, KK5 and gauge fields on D-brane are new ingredients, because those objects have not been considered in 
\cite{bio_WB,bio_KR,bio_BBST1,bio_BW,bio_W2,bio_BP1,bio_BP2, bio_P2, bio_KS, bio_CWB, bio_C1, bio_BC, bio_BBC, bio_ET, bio_CW, bio_DFB}. 
The reason why the moduli fixing has been difficult in the brane gas model so far 
is that we do not deal with the effect of the gauge fields on the D-branes, 
the world volume action of NS5-branes and KK5-monopoles. 

The key idea of introducing NS5-branes and KK5-monopoles is 
the difference of the dilaton dependence between D-brane and NS5-brane/KK5-monopole. 
In the string frame, the dilaton dependence of D-brane is $e^{-\phi}$, on the other hand, 
NS5-branes and KK5-monopoles have the dependence of $e^{-2\phi}$. 
This fact implies that NS5-branes and KK5-monopoles generate the stronger pressure for the dilaton than D-branes. 

A brane gas model whose moduli fields are fixed can be mapped to 
a dual brane gas model with fixed moduli fields by the T-duality transformations. 
Once moduli fields are fixed,  
we are able to obtain dual type IIA or IIB brane gas models by T-duality. 
Then, the fixed values of the moduli fields are not necessarily self-dual. 
This fact is different from String gas model \cite{bio_BV} where moduli fields are fixed 
at self-dual radius of T-dual without the dilaton field.  

In our model, the problem on the selection of vacua remains. 
At this stage, the brane configurations, the number of branes and the initial values of gauge fields 
are given by hand. A lot of initial conditions give the various minimum of the potential. 
However, once the initial condition is given, the moduli fields can be fixed. 

It is interesting to classify the cases where the moduli fields are fixed in our model. 
This analysis may lead to the moduli fixing which includes non-diagonal components of the metric of $T^{6}$. 

 \newpage
The toroidal compactification with fixed moduli fields is a first and significant step to obtain the realistic compactification with 
$K3 \times T^{2}$ or Calabi-Yau threefold \cite{bio_E}. Our analysis of this paper includes the configuration where
 D2-branes wrap around two-cycles in type IIA theory and D3-branes wrap around three cycles in type IIB theories. 
Therefore, it is expected that the similar analysis can be made even for the models with Calabi-Yau compactifications.
The moduli fixing procedure discussed in our paper may also be applicable to a heterotic brane gas model 
with $T^{6}$ compactification which is expected to be dual to a type IIA with $K3 \times T^{2}$ compactification \cite{bio_HT, bio_D, bio_W}.

An application of our model to the accelerating universe \cite{bio_TW} is also interesting.
The method used in \cite{bio_SS} may be useful for the analytic investigation. 

\section{acknowledgments}
We would like to thank the Yukawa Institute for Theoretical Physics at Kyoto University. 
Discussions during the YITP-W-07-05 on ``String theory and field theory - frontier of quantum and space-time'' 
and during the YITP-W-07-10 on "KIAS-YITP Joint Workshop "String Phenomenology and Cosmology" were useful to complete this work. 
We are grateful to thank Kenji Hotta, Katsumi Itoh, Testuji Kimura, Nobuyoshi Ohta, Yuji Satoh, Makoto Tanabe , Seiji Terashima and 
Kunihito Uzawa for valuable comments.

\end{document}